\documentstyle[12pt,epsf,epsfig,wrapfig]{article}
\newcommand{\rar}{\rightarrow}
\newcommand{\lar}{\leftarrow}
\newcommand{\pdup}{p_\uparrow}
\newcommand{\ddup}{d_\uparrow}

\newcommand{\pimd}{\pi^- + \ddup \rar \pi^0 + X}

\newcommand{\pime}{\pi^- + \pdup \rar \pi^0 + n}
\newcommand{\pimw}{\pi^- + \pdup \rar \omega (782) + n}
\newcommand{\pimetap}{\pi^- + \pdup \rar \eta \prime (958) + n}

\def\Journal#1#2#3#4{#1 {\bf #2}, #3 (#4)}

\begin{document}

\begin{center}
{\bfseries NEW POLARIZATION PROGRAM AT U70 (SPASCHARM PROJECT)
}

\vskip 5mm

\underline{A.N.~Vasiliev$^1$}, V.V.~Mochalov$^{1 \dag}$, S.I.~Alekhin$^1$, 
N.A.~Bazhanov$^2$, N.I.~Belikov$^1$, A.A.~Belyaev$^3$, 
N.S.~Borisov$^2$, B.V.~Chujko$^1$, Y.M.~Goncharenko$^1$, 
V.N.~Grishin$^1$, A.M.~Davidenko$^1$, A.A.~Derevschikov$^1$, 
V.A.~Kachanov$^1$, V.Y.~Kharlov$^1$, A.S.~Kozhin$^1$, 
D.A.~Konstantinov$^1$, V.G.~Kolomiets$^2$, V.A.~Kormilitsin$^1$,
V.I.~Kravtsov$^1$, A.B.~Lazarev$^2$, A.K.~Likhoded$^1$, 
A.V.~Luchinsky$^1$, A.A.~Lukhanin$^3$,Yu.A.~Matulenko$^1$, 
Yu.M.~Melnick$^1$, A.P.~Meschanin$^1$, N.G.~Minaev$^1$, 
D.A.~Morozov$^1$, A.B.~Neganov$^2$, L.V.~Nogach$^1$, 
S.B.~Nurushev$^{1}$, Yu.A.~Plis$^2$, A.F.~Prudkoglyad$^1$, 
A.V.~Ryazantsev$^1$, P.A.~Semenov$^1$, O.N.~Shchevelev$^2$, 
S.R.~Slabospitsky$^1$, L.F.~Soloviev$^1$, M.N.Ukhanov~, 
Yu.A.~Usov$^2$, A.V.~Uzunian$^1$, A.S.~Vovenko$^1$, 
A.E.~Yakutin$^1$ \\

\vskip 5mm
{\small 
(1) {\it Institute for High Energy Physics, Protvino, Russia}\\
(2) {\it Joint Institute for Nuclear Research, Dubna, Russia}\\
(3) {\it Kharkov Institute of Physics and Technology, Ukraine}\\
$\dag$ {\it E-mail: mochalov@ihep.ru}}

\end{center}

\vskip 5mm

\begin{abstract}
The new polarization program SPASCHARM is being prepared in Protvino. 
The program has two stages. The first stage is dedicated to single-spin 
asymmetries in the production of miscellaneous light resonances with 
the use of 34~GeV $\pi^-$-beam. Inclusive and exclusive reactions will 
be studied simultaneously. The second stage is dedicated to single-spin 
and double-spin asymmetries in charmonium production with the use 
of 70~GeV polarized proton beam which will allow us to understand 
charmonium hadronic production mechanism and make gluon polarization 
$\Delta g(x)$ extraction at large $x$. 
\end{abstract}

\vskip 8mm

\section*{Introduction}
A possibility to accelerate high-intensive polarized proton beam up to 
70~GeV at the IHEP U70 accelerator, extract it from the main ring and 
deliver to several experimental setups was intensively studied last 
time in 2005 and Spring of 2006 in Protvino~\cite{spin05}-\cite{praga06}. 
We proposed to study a wealth of single- and double-spin observables 
in various reactions using longitudinally and transversely polarized 
proton beams at U70. Unfortunately the proposal stuck in the Ministry 
of Education and Science in Summer 2006. But we believe that a 
possibility to push the proposal still exists.  

The main goal of the SPASCHARM project is to study spin structure of 
the proton, starting with determination of gluon contribution into 
spin of the proton at large Bjorken $x$ through study of spin effects 
in charmonium production. High sensibility to gluon content of the 
interacting particles is one of the main features of charmonia 
production in hadronic interactions. In case of collision of two 
longitudinally polarized protons it is used to define gluon polarization 
$\Delta G/G$ in the proton.  A polarized proton beam is needed to make 
this study. We plan to have it at the second stage of the experiment 
after the measurements of single-spin asymmetries already in charmonia 
production have been carried out.

The project has a first stage when unpolarized beams will be used. 
The first stage is an experiment to study single-spin asymmetries $A_N$ 
of light resonances consisting of $u$-, $d$- and $s$-valence quarks. 
Transverse single-spin asymmetries are very well known for a long time. 
In the Standard Model QCD at leading twist level all $A_N=0$. But the 
experiments show very big $A_N$ in the confinement region. Therefore 
$A_N$ is very sensitive to the effects outside the $SM$. The known 
theoretical approaches (Sivers and Collins effects, twist-3 effect, 
etc.) try to reconcile theory and experiment. To discriminate the 
existing theoretical approaches and to stimulate to develop the new 
ones, a systematic study of $A_N$ for a big number of miscellaneous 
inclusive and exclusive reactions is needed, especially in the 
confinement region, which is the most unclear for theory.  
To make this systematic study is the main goal of the first stage 
of the SPASCHARM project. The first stage will be finalized by the 
measurements of $A_N$ in charmonia production. This will finally 
prepare the experimental setup to the second stage of the project 
where only one new thing will be needed - namely a polarized proton 
beam from U70. 

This paper is organized as follows. First we will describe the second 
stage of the experiment dedicated to spin effects in charmonia 
production with the use of polarized proton beam from U70. 
After that we will describe the first stage dedicated to spin effects 
in light resonance production.

\section*{Charmonia production in polarized $p_{\rar}p_{\rar}$ 
interactions}

\begin{wrapfigure}{R}{6.5cm}
\hspace*{0.3cm}
\mbox{\epsfig{figure=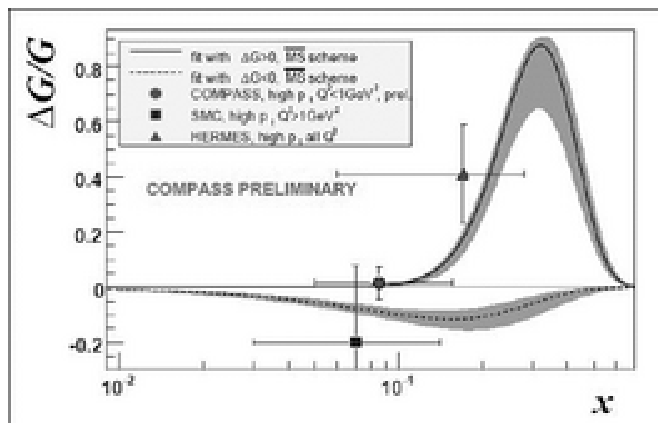,width=6.2cm}}
\hspace*{0.3cm}
\begin{minipage}[t]{6.cm}
%\vspace*{-0.2cm}
{\small{\bf Figure 1.} 
{The solution of $\Delta G/G$ from experiment COMPASS~\cite{santos}.}}
\end{minipage}
\end{wrapfigure}

At present only 30\% of the longitudinally polarized proton spin is 
described by quark's spin. The other 70\% of the proton spin may be 
explained by gluon and/or orbital momentum contributions. Experiments with 
polarized lepton beams at CERN, HERA, SLAC have been measuring mainly 
quark polarization over last twenty years. COMPASS and HERMES have 
tried to measure gluon polarization at small $x$, up to 0.1-0.15. 
The RHIC experiments STAR and PHENIX have begun to measure gluon 
polarization at very low $x$ values (about 0.01) whereas gluon polarization 
has to be measured in the whole $x$ range. So in spite of many years of 
experiments, a detailed decomposition of the spin of the proton remains 
elusive - new experimental data on $\Delta g(x, Q^2)$, especially at 
large $x$ are badly needed. We propose to simultaneously measure 
the double-spin asymmetry $A_{LL}$ for inclusive $\chi_{c2}, \chi_{c1}$ 
and $J/\Psi$  by utilizing the 70~GeV/c longitudinally polarized-proton 
beam on a longitudinally polarized target. Our goal is to obtain 
besides the quark-spin information also the gluon-spin information 
from these three processes in order to determine what portion of the 
proton spin is carried by gluons. Better understanding of charmonium 
production at U70 energies is needed -- for this pion and proton beams 
will be used to produce charmonium. Gluon contribution into the proton 
spin as well as strange quarks and orbital momentum contributions -- 
worldwide studies at HERMES, COMPASS, RHIC, JLAB and SLAC. 
We propose a new experiment in this field -- it should be complimentary 
to the existing experiments. It will give new data at large $x$ for 
Global analysis. One can see from Fig.1 that the biggest gluon 
polarization is anticipated near $x=0.3$. SPASCHARM will measure gluon 
polarization in the region of $x$ between 0.3 and 0.6.

\begin{wrapfigure}{R}{4.2cm}
\hspace*{0.2cm}
\mbox{\epsfig{figure=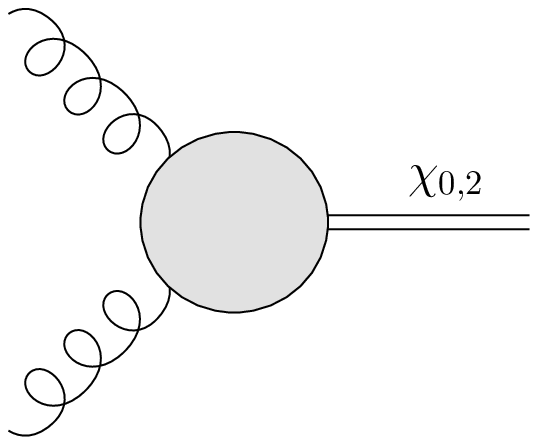,width=3.8cm}}
\hspace*{0.2cm}
\begin{minipage}[t]{4.cm}
%\vspace*{-0.2cm}
{\small{\bf Figure 2.} 
{Gluon fusion ($\alpha_S^2, p_T=0$).}}
\end{minipage}
\end{wrapfigure}

Information about gluon polarization might be obtained through 
simultaneous measurements of $A_{LL}$ in inclusive production of 
$\chi_{c2}$ and $J/\Psi$.  This experiment was proposed at 
Fermilab (P838) at 200 GeV as a continuation of E704~\cite{fnalprop}. 
The Fermilab's PAC pointed out that physics was very interesting, 
but an intensity of the polarized proton beam  from $\Lambda$-hyperon 
decays was small -- the statistics would not be enough. 
The experiment was not approved. In our new proposal for U70 we 
expect to have up to $4\cdot 10^8$~p/min instead of 
$2.7 \cdot 10^7$~p/min in P838 which is a factor of 15 more.

\begin{figure}[h!]
\begin{center}
\epsfig{figure=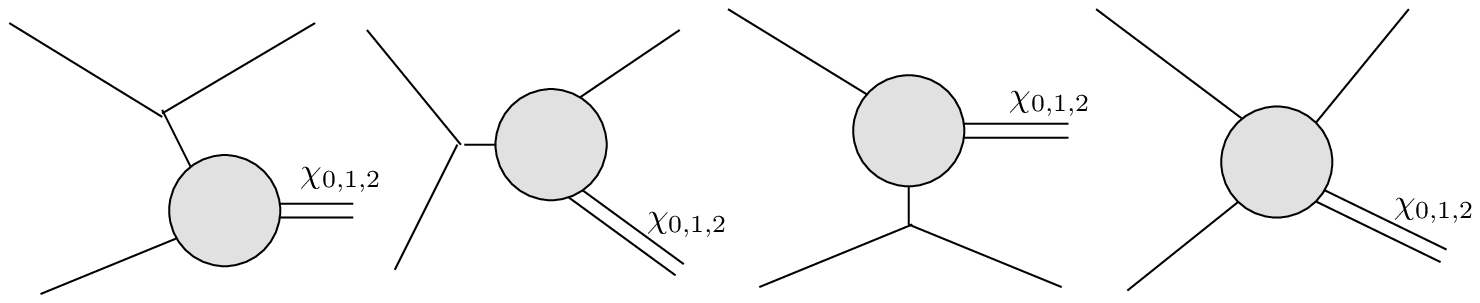,width=0.9\textwidth} 
\vskip -3mm
{\small{\bf Figure 3.} 
{Gluon fusion ($\alpha_S^3$).}}
\end{center}
\end{figure}

\begin{figure}[b!]
\begin{center}
\begin{tabular}{p{0.55\textwidth}p{0.07\textwidth}p{0.31\textwidth}}
\mbox{\epsfig{figure=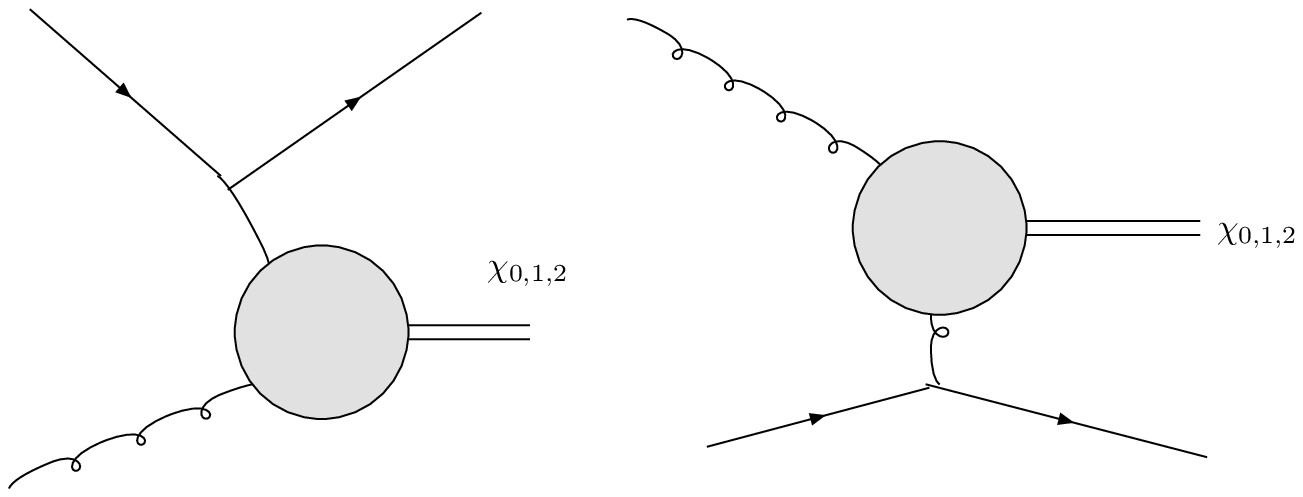,width=0.52\textwidth}} & & 
\mbox{\epsfig{figure=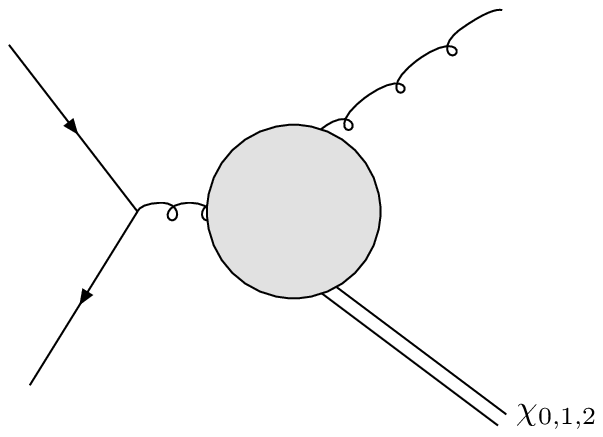,width=0.29\textwidth}}\\ 
{\small{\bf Figure 4.} Quark-gluon interaction ($\alpha_S^3$).} & &
{\small{\bf Figure 5.} Quark-antiquark annihilation ($\alpha_S^3$).}\\
\end{tabular}
\end{center}
\end{figure}

The hadronic production of the $\chi$ states involves three 
parton fusion diagrams~\cite{likhoded}:
\begin{enumerate}
\item{\vspace*{-0.3cm} gluon fusion (Fig.2-3)};
\item{\vspace*{-0.3cm} quark-glion interaction (Fig.4);}
\item{\vspace*{-0.3cm} quark-antiquark annihilation (Fig.5).}
\end{enumerate}

Estimate made by one of our authors (S.A.Alekhin) has shown that at 
70~GeV the contributions of gluon-gluon fusion and quark-antiquark 
annihilation to produce charmonium with a mass of 3.5~GeV in 
$pp$-interactions are comparable.
 
The goal of the proposed experiment is to measure double-spin 
asymmetry $A_{LL}$  with the use of longitudinally polarized beam 
and target in the process: 
\begin{equation}
p_{\rar}+p_{\rar} \rar \chi_{c2}(J/\Psi)+X, (\chi_{c2} \rar J/\Psi +\gamma).
\end{equation}

$J/\Psi$ will be registered mainly via $\mu^+\mu^-$ decay due to 
bremmstrahlung in $e^+ e^-$ decay mode. The charmonia states under study 
are $J/\Psi$~(3096, $J^{PC} = 1^{--}$), $\chi_{c1}$~(3510, $J^{PC}=1^{++}$) 
and $\chi_{c2}$~(3555, $J^{PC}=2^{++}$). The measured experimental asymmetry 
is given by 
\begin{equation}
A_{LL}= \frac{1}{P_B \cdot P_T^{eff}}\cdot \frac{I^{++}-I^{+-}}{I^{++}+I^{+-}},
\end{equation}

\noindent where $P_B$ is the beam polarization, $P_T^{eff}$ -- effective 
target polarization, $I^{++}, I^{+-}$ are the number of events normalized 
to the incident beam. The helicity states (++) and (+-) correspond to 
($\lar \rar$) and ($\rar \rar$) states respectively, where arrows 
indicate the beam and target spin direction in the laboratory system.

Theoretical predictions of $A_{LL}$ mainly depend on two assumptions:
\begin{itemize}
\item
\vspace*{-0.3cm} gluon polarization $\Delta G/G$  and 
\item
\vspace*{-0.3cm} charmonium production mechanism which defines   
$\hat{A}_{LL}$  at the parton level (in parton-parton interaction). 
\end{itemize}

\begin{figure}[h!]
\begin{center}
\epsfig{figure=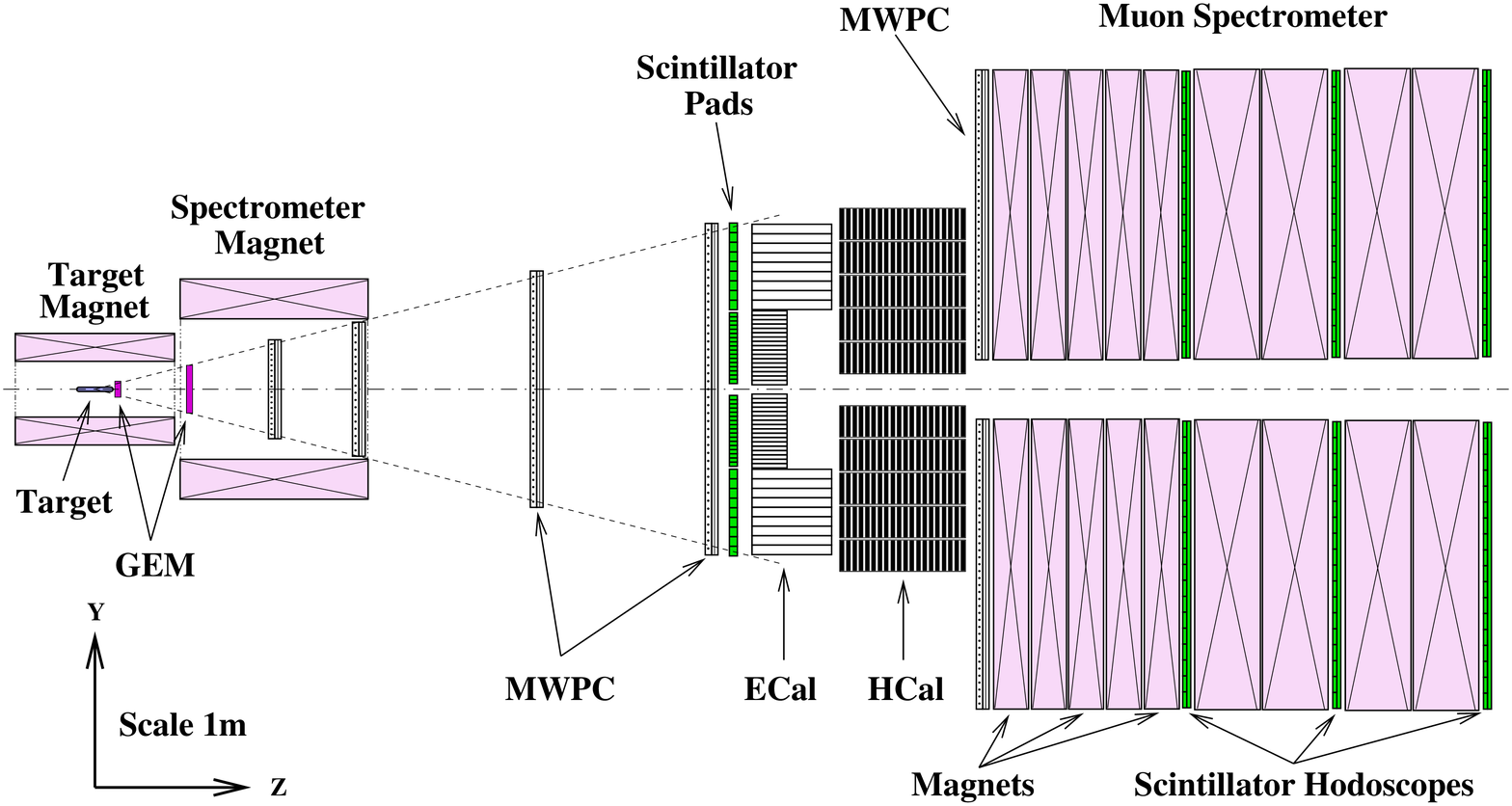,width=0.8\textwidth} 
\vskip -2mm
{\small{\bf Figure 6.} 
SPASCHARM Experimental Setup}
\end{center}
\end{figure}

The experimental setup SPASCHARM is presented in Fig.6. 
It is an open geometry experiment. The main parts of the setup are as 
follows: 
\begin{itemize}
\item
\vspace*{-0.3cm} 
wide aperture spectrometer with GEM, drift chambers and proportional 
chambers; 
\item
\vspace*{-0.3cm} electromagnetic calorimeter and 
\item
\vspace*{-0.3cm} muon detector. 
\end{itemize}

The central part of the calorimeter (1~m$^2$) will consist of lead 
tungstate blocks. It is critically needed to detect very precisely 
$\gamma$-quanta fro $\chi$-decays to separate $\chi_{c1}$ and $\chi_{c2}$ through
high precision energy resolution of the calorimeter.
 The $x_F$ distribution of $\chi_{c2}$~(3555) detected
by the setup at a beam energy of 70~GeV is presented in Fig.~7.

\begin{figure}[t!]
\begin{center}
\begin{tabular}{p{0.43\textwidth}p{0.04\textwidth}p{0.43\textwidth}}
\mbox{\epsfig{figure=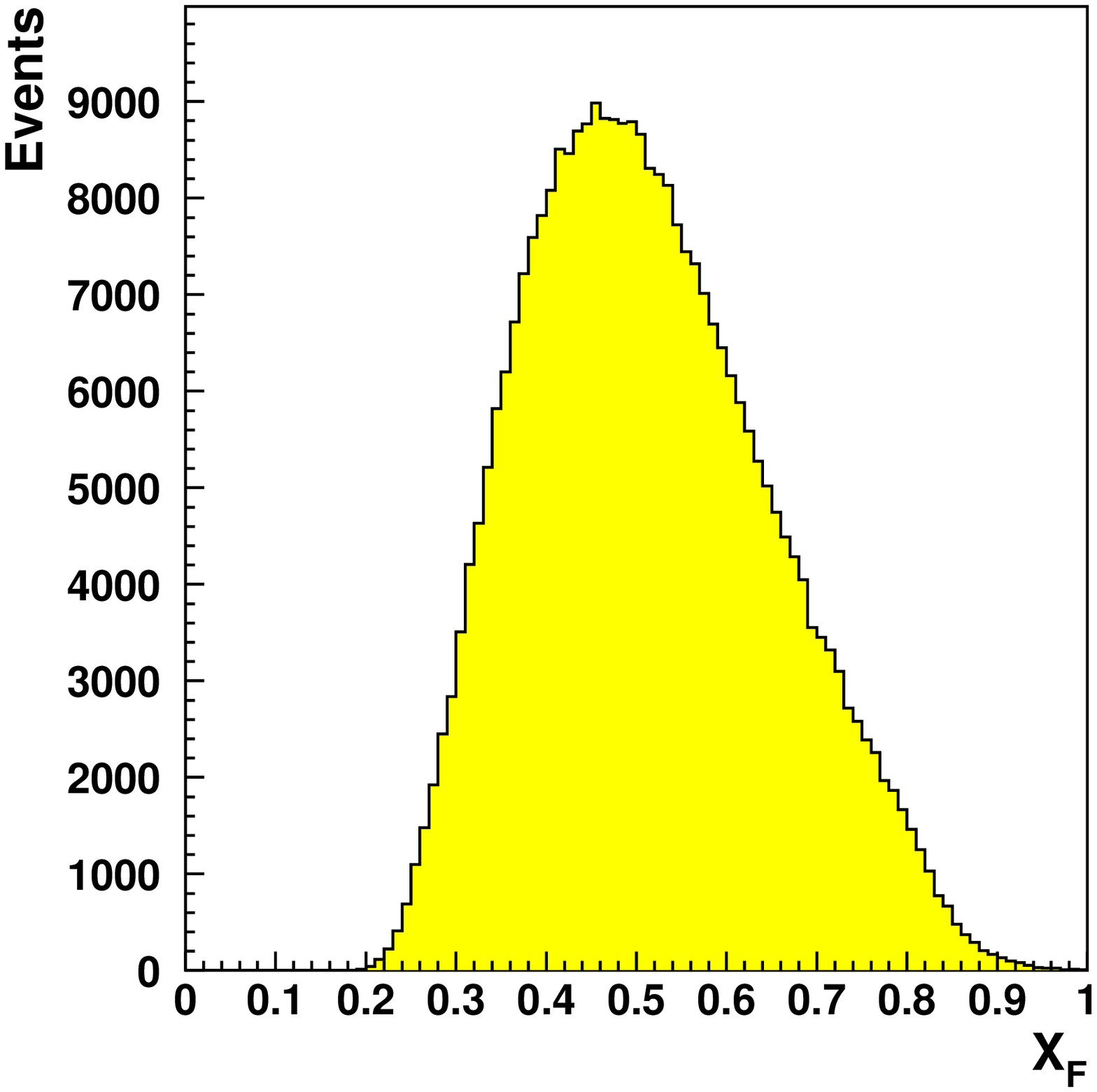,width=0.4\textwidth}} & & 
\mbox{\epsfig{figure=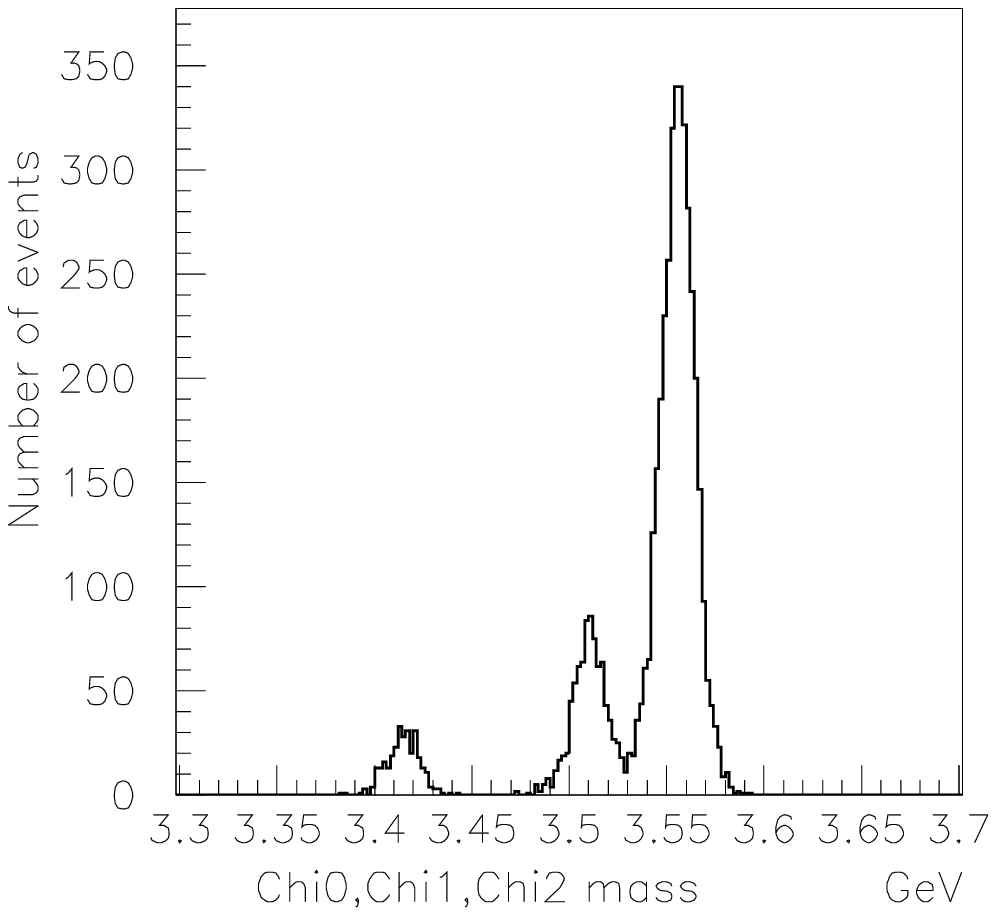,width=0.4\textwidth}}\\ 
{\small{\bf Figure 7.} The $x_F$ distribution of 
$\chi_{c2}$~(3555) detected by the setup at a beam 
energy of 70~GeV} & &
{\small{\bf Figure 8.} The reconstructed masses of 
$\chi_{c0}$~(3410), $\chi_{c1}$~(3510) and $\chi_{c2}$~(3555) 
as a result of Monte-Carlo simulations for the SPASCHARM 
experimental setup.}\\
\end{tabular}
\end{center}
\end{figure}

The principal point for this experiment is a separation of the two 
charmonia states with the spins equal to 1 and 2, namely 
$\chi_{c1}$~(3510) and $\chi_{c2}$~(3555). The Monte-Carlo simulations 
for 70~GeV have been made.  The reconstructed masses of 
$\chi_{c0}$(3410), $\chi_{c1}$~(3510) and $\chi_{c2}$~(3555) are 
presented in Fig.8.  The $J/\Psi$ (in $\mu \mu$-decay mode) 
4-momentum is taken as a result of $1C$-fit. For charged particles 
$\Delta p/p$= 0.004 at 10 GeV/c.  For $\gamma$-quanta 
$\sigma (E)/E$ was taken as $2.5\%/\sqrt{E}$. We can see that the two 
states of interest are well separated. 

The SPASCHARM experiment plans to have 25000 electronic channels 
(7000 ADC, 2000 TDC and 16 000 registers). The trigger for interaction 
in the target will be the only hardware trigger. Information from the 
interaction will be digitized in each sub-detector, pre-processed and 
buffered for further processing.  A high level trigger selection will 
occur in compute nodes which access the buffers via a high bandwidth 
network fabric.  The experiment plans to operate at 
interaction rates of the order of 2~MHz. With pre-processing on the 
detector electronics for a substantial reduction of the data volume, 
typical event sizes are in the range of 2 to 4 ~kB. This amounts to 
total raw data rates in the order of 3~GB/s. 

Our estimate has shown us that we expect to get a precision of 
$\sigma (A_{LL})$ = 0.07 for $\chi_{c2}$ and 0.025 for $J/\Psi$ at 
$x=0.3$ for 100 days of data taking. 

With the use of polarized proton beam at SPASCHARM a precision 
measurement of single-spin asymmetry in inclusive production of 
miscellaneous resonances in the transverse polarized beam 
fragmentation region in a wide ($x_F, p_T$)-region will be worthwhile. 
Also it will be possible to measure transversity in Drell-Yan muon 
(electron) pairs. 

\section*{Single-spin asymmetries in light resonance production}

Before the polarized proton beam will be accelerated at U70 we can 
make single-spin measurements of miscellaneous inclusive and exclusive 
reactions with unpolarized beams, such as pions, kaons and protons, 
existing at the beam channel 14 of the Protvino accelerator.  Why do 
we need to measure $A_N$ in a big variety of inclusive and exclusive 
reactions? In the Standard Model QCD at leading twist level 
all $A_N$=0. But the experiments show very big $A_N$ in the confinement
 region. Therefore $A_N$ is very sensitive to the effects outside 
the $SM$. The known theoretical approaches (Sivers and Collins 
effects, twist-3 effect, etc.) try to reconcile theory and 
experiment. To discriminate the existing theoretical approaches and 
to stimulate to develop the new ones, a systematic study of $A_N$ 
for a big number of miscellaneous inclusive and exclusive reactions 
is needed, especially in the confinement region, which is the most 
unclear for theory.  To make this systematic study is the main goal 
of the first stage of the SPASCHARM project.

\begin{wrapfigure}{R}{8.5cm}
\vspace*{-1.cm}
\hspace*{0.2cm}
\mbox{\epsfig{figure=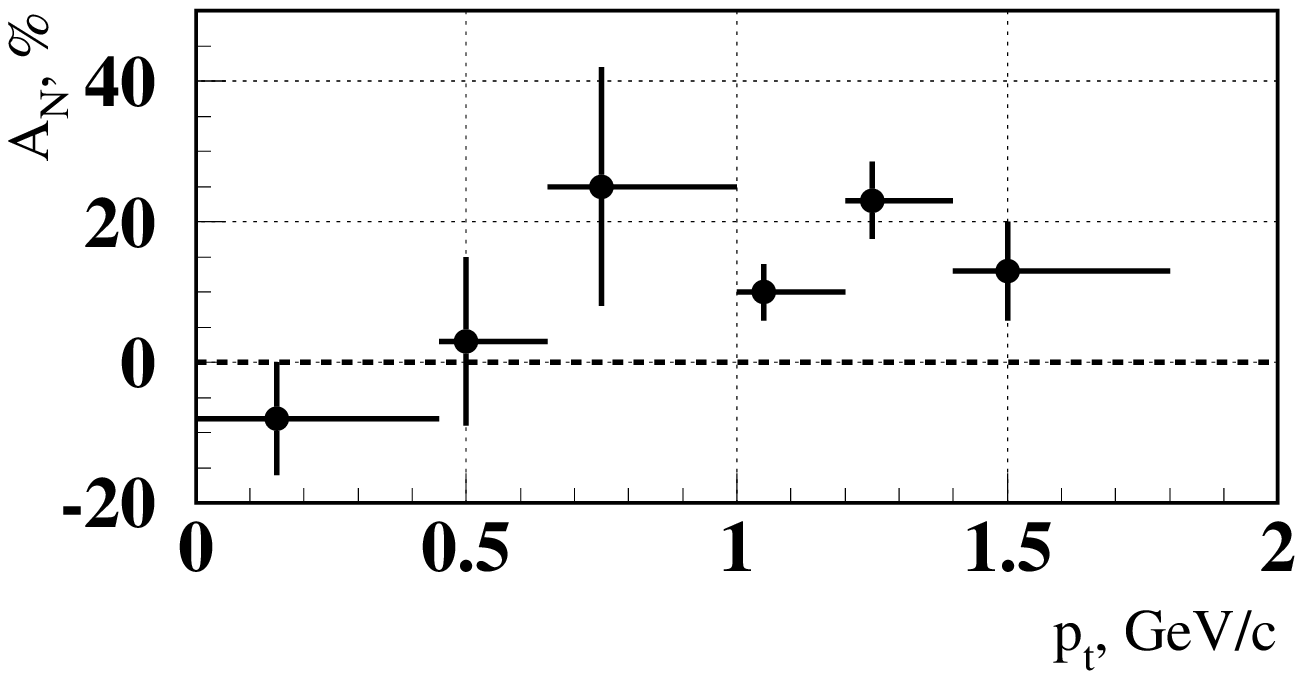,width=7.9cm}}
%\hspace*{0.1cm}
\begin{minipage}[t]{8.3cm}
\vspace*{-0.9cm}
{\small{\bf Figure 9.} 
{The $p_T$-dependence of single-spin asymmetry 
$A_N$ in the inclusive  reaction $\pimd$ at 40~GeV/c at $x_F>0.7$. 
The average value of $A_N$ is $(16 \pm 5)$\% near $p_T$ equal 
to 1~GeV/c.}}
\end{minipage}
\hspace*{0.2cm}
\mbox{\epsfig{figure=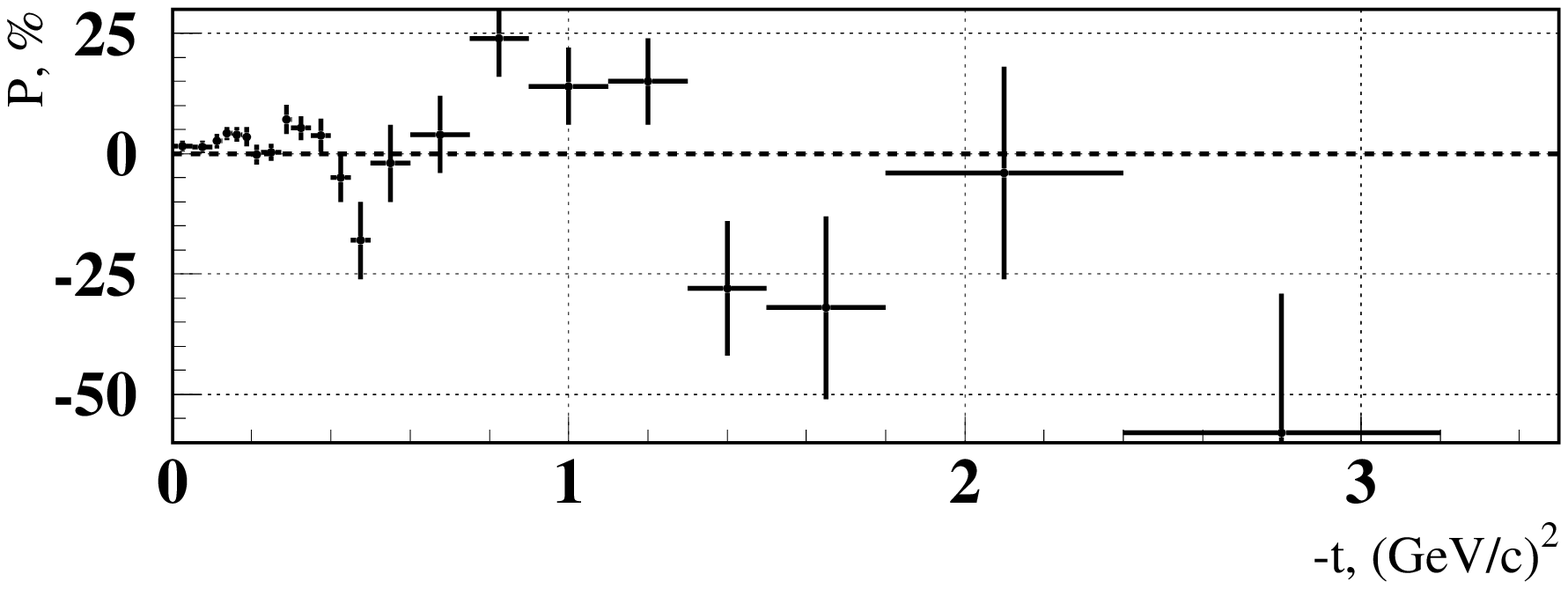,width=8.3cm}}
%\hspace*{0.2cm}
\begin{minipage}[t]{8.1cm}
\vspace*{-0.7cm}
{\small{\bf Figure 10.} 
{The $t$-dependence of $A_N$ in the exclusive 
reaction $\pime$ at 40~GeV/c. The average value of $A_N$ is 
$(18\pm 5)$\% near $t$ equal to 1~(GeV/c)$^2$.}}
\end{minipage}
\end{wrapfigure}

It would be interesting to measure single-spin asymmetries in 
inclusive production of light resonances even in the unpolarized beam 
fragmentation region, but at big values of transverse momentum $p_T$, 
close to the boundary of phase space. In Fig.9 the single-spin 
asymmetry $A_N$ in the inclusive reaction 
$\pimd$ at 40~GeV/c at $x_F>$0.7 is presented~\cite{pidforw}. We see that $A_N$ 
is zero at small $p_T$ and about 15\% at $p_T$ near 1~GeV/c and bigger. 
When $x_F$ goes to 1, any inclusive reaction transfers into the proper 
exclusive reaction. In Fig.10 the single-spin asymmetry $A_N$ in the 
exclusive reaction $\pime$ at 40~GeV/c is presented~\cite{pin}. We see that 
$A_N$ is also about 15\% near  $-t$ equal to 1~(GeV/c)$^2$, that is 
equivalent to $p_T$ near 1~GeV/c. So asymmetries in the both inclusive 
and exclusive $\pi^0$-production at 40~GeV pion beam are equal each 
other (also it seems that asymmetries on polarized protons and 
neutrons are the same). It should be the case for other light 
resonances.

For the first stage of the experiment two multi-channel threshold 
Cherenkov counters will be added to the setup to distinguish between 
pions and kaons. They are of 1.5~m and 3~m long and will be placed 
between the end of the magnet and the calorimeter. They will be 
filled by freon and by air correspondingly, both at atmospheric 
pressure. Lead tungstate in the calorimeter is not needed for the 
first stage, lead glass with moderate energy resolution will be 
enough to detect light resonances. An acceptance of the whole setup 
will be decreased, however it will still be significant to detect 
light resonances. Due to very fast DAQ (practically without dead time)
inclusive and exclusive reactions will be studied simultaneously.

%As an example, some typical mass spectra obtained after reconstruction 
%of the Monte-Carlo simulated events at 70~GeV proton beam are 
%presented in Fig.11-13. We see pretty clean mass peaks of the proper 
%resonances in these spectra.  

There are some advantages of the new experiment. Exclusive and 
inclusive reactions were studied either in neutral decay modes or in 
charged decay modes in the previous experiments. We propose to measure 
the both modes simultaneously and therefore we expect a significant 
increase in statistics. Addition of new detectors (GEM, MDC, high 
quality EMC etc.) compare to the previous experiments might bring us 
to discovery of "new channels" (exotic glueballs, hybrids, etc). 
Extremely high-speed DAQ will allow to detect inclusive and exclusive 
reactions simultaneously. Partial wave analysis of a huge statistics 
on polarized target will raise a robustness of the results on rare 
resonances. The setup has $2\pi$-acceptance on azimuthal angle $\phi$ 
and therefore the systematic errors in single-spin asymmetries will 
be negligible.

\begin{wrapfigure}{R}{9.5cm}
\vspace*{-1.cm}
\hspace*{0.2cm}
\mbox{\epsfig{figure=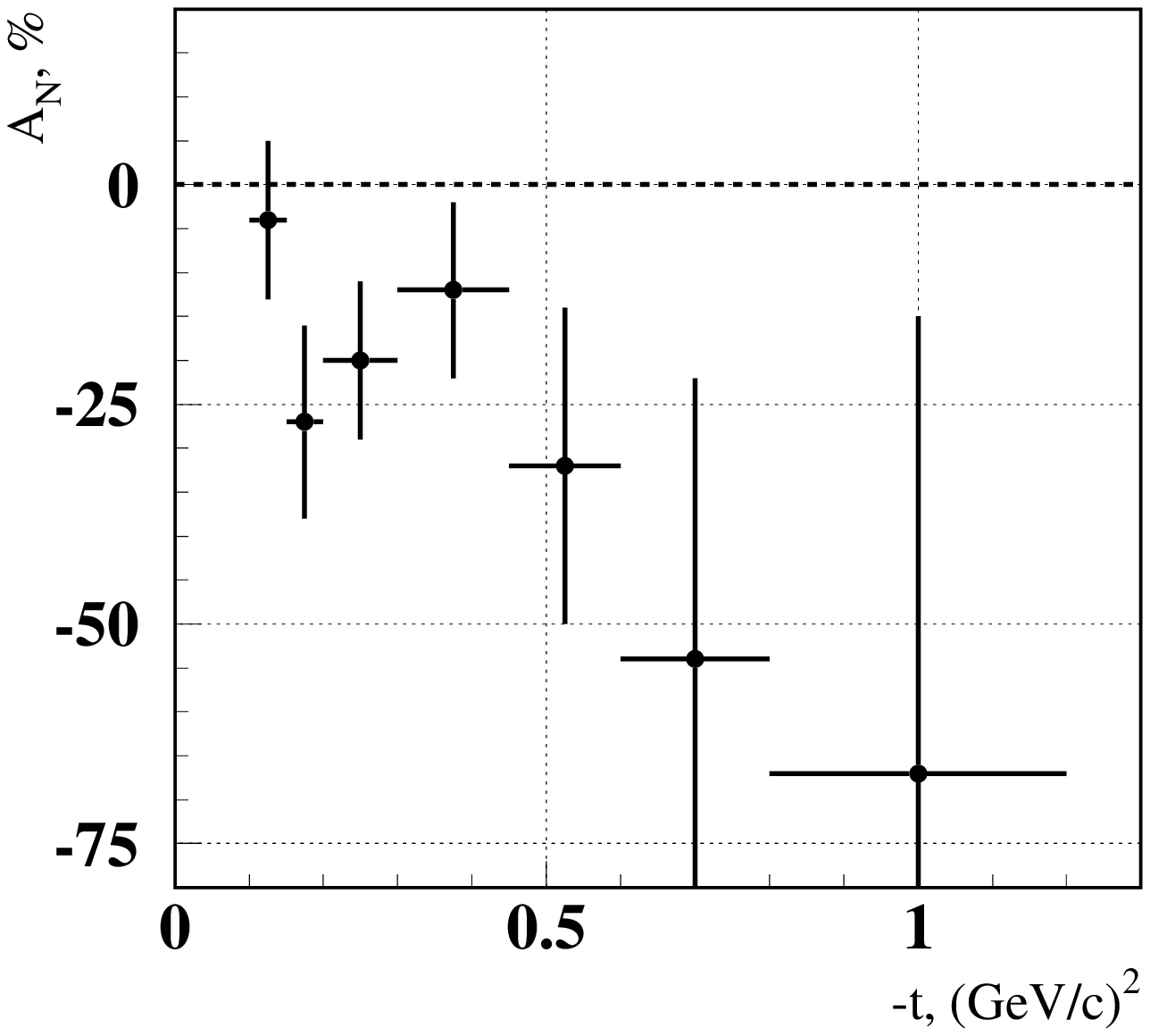,width=9.2cm, height=5.8cm}}
%\hspace*{0.1cm}
\begin{minipage}[t]{9.4cm}
\vspace*{-0.9cm}
{\small{\bf Figure 11.} 
{$A_N^{\pimw}$ at 40~GeV~\cite{piomega}. The $\omega$~(782) has 
been detected in $\pi^0 \gamma$  decay mode with 8\% branching.  
33,000 events on polarized target were collected. 
Solid angle was twice less than in the SPASCHARM setup for the 
first stage. By using two decay modes ($\pi^+\pi^-\pi^0$ with 89\% 
branching and $\pi^0 \gamma$), statistics can be increased in 20 times. 
Errors in the first four points would be 
2\% rather than 10\% now.}}
\end{minipage}
\hspace*{0.2cm}
\mbox{\epsfig{figure=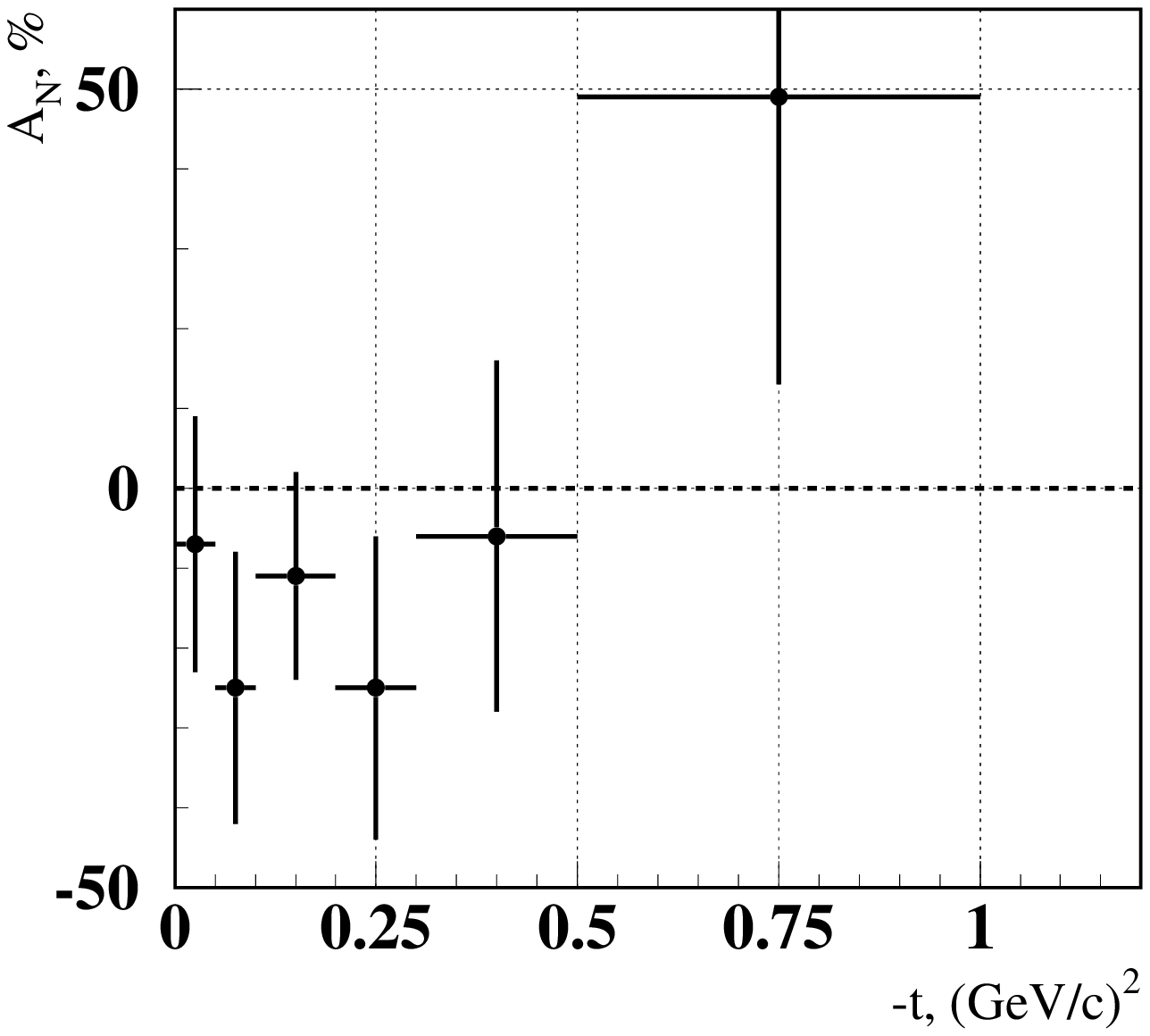,width=9.2cm,height=5.8cm}}
%\hspace*{0.2cm}
\begin{minipage}[t]{9.4cm}
\vspace*{-0.9cm}
{\small{\bf Figure 12.} 
{$A_N^{\pimetap}$ at 40~GeV~\cite{pieta}.  The $\eta \prime$~(958) 
has been detected in $\gamma \gamma$ decay mode with 2\% branching.  
11,000 events on polarized target were collected. Solid angle was 
about the same as in the SPASCHARM setup for the 
first stage. By using two additional decay modes ($\pi^+\pi^-\eta$ and 
$\pi^+\pi^-\gamma$ with branchings of 45\% and 30\%), statistics can be 
increased in 20 times. Errors in the first 
three points would be 3-4\% rather than 13-17\% now.
}}
\end{minipage}
\end{wrapfigure}

One can see the advantage of proposed new measurements in sense of 
significant increase in statistics in a couple exclusive reactions 
in Fig.11 and Fig.12. The details are in the Figure captions. 
        
For the MC simulations, two options of the setup were considered 
with two distances from the center of the polarized target to the 
beam downstream end of the last Cherenkov counter - "7 meters" and 
"4 meters". Variant "4 meters" has one Cherenkov counter in the 
setup. $\pi$-mesons will be identified in the momentum region of 
3-23 GeV/c. 
%No $\pi$ /K-separation. 
Acceptance for "usual" (non-strange) resonances is huge (3 times 
bigger than for "7 m"). We request a beam time of 30 days. 
Variant "7 meters" has two Cherenkov counters in the setup and 
allows $\pi /K$-separation in the momentum region of 3-23 GeV/c.  
We request a beam time of 70 days. The expected accuracies of $A_N$ in 
several inclusive reactions for the summing 100 days at beam in the 
kinematical region of   $x_F= 0.5-1.0$ and $p_T=0.5-2.5$~GeV/c are 
the following for different reactions:\\
$\sigma (A_N^{\pi^- + \pdup \rar \omega + X})$ = 0.3-3\%;\\
$\sigma (A_N^{\pi^- + \pdup \rar \rho + X})$ = 0.2-2.5\%;\\
$\sigma (A_N^{\pi^- + \pdup \rar \eta \prime + X})$ = 0.3-4\%;\\
$\sigma (A_N^{\pi^- + \pdup \rar f_2 + X})$ = 0.1-1\%;\\
$\sigma (A_N^{\pi^- + \pdup \rar \phi + X})$ = 3.-10.\%;\\
$\sigma (A_N^{\pi^- + \pdup \rar K^{*0} + X})$ = 0.6-10\%.

\section*{Conclusion}

The new polarization program SPASCHARM is being prepared in Protvino.
The program has two stages. The first stage (to be started in 2011) is 
dedicated to single-spin asymmetries in the production of 
miscellaneous light resonances with the use of 34~GeV $\pi^-$-beam. 
Inclusive and exclusive reactions will be studied simultaneously. 
The errors in the exclusive reactions with big asymmetries are 
expected to be several times less than now. The brand new data for 
inclusive reactions will be obtained. All the new data will much 
better help us to understand spin dependence of strong interaction 
in the most difficult from the theory point of view kinematical 
region, namely in the quark confinement region. 

The second stage (to be started in 2015) is dedicated to single-spin 
and double-spin asymmetries in charmonium production with the 
use of 70~GeV polarized proton beam  which will allow us to 
understand  charmonium hadronic production mechanism and make 
$\Delta g(x)$ extraction at large $x$. The results on $\Delta g(x)$ 
at large $x$ will be unique and will be complementary to those 
which exist and might be obtained at COMPASS, HERMES, RHIC and JLAB 
at smaller $x$. The global analysis with the use of the new large 
$x$ data on $\Delta g(x)$ will significantly improve our knowledge 
of the gluon polarization integral $\Delta G$.

This work has been partially supported by the RFBR grant 06-02-16119.

\end{document}